\documentclass[sn-mathphys-num,iicol]{sn-jnl}

\usepackage{graphicx}%
\usepackage{multirow}%
\usepackage{amsmath,amssymb,amsfonts}%
\usepackage{amsthm}%
\usepackage{mathrsfs}%
\usepackage{xcolor}%
\usepackage{textcomp}%
\usepackage{manyfoot}%
\usepackage{booktabs}%
\usepackage{algorithm}%
\usepackage{algorithmicx}%
\usepackage{algpseudocode}%
\usepackage{listings}%
\usepackage{slashed, mathtools}%
\usepackage{cuted}%
\begin{document}

\title{Ginzburg-Landau approach to the Gross-Neveu model: success and failure}


%\begin{center}
%\author{Lalita Choudhary, Anees Ahmed \\
%Department of Physics, MNIT Jaipur, India \\
%\small{email:\emph{ \href{mailto:anees.phy@mnit.ac.in}{anees.phy@mnit.ac.in}}}}
%\end{center}

\author[]{\fnm{Lalita} \sur{Choudhary}}\email{2022rpy9089@mnit.ac.in}

\author*[ ]{\fnm{Anees} \sur{Ahmed}}\email{anees.phy@mnit.ac.in}

\affil[]{\orgdiv{Department of Physics}, \orgname{MNIT Jaipur}, \orgaddress{\postcode{302017}, \state{Rajasthan}, \country{India}}}

\abstract{The phase diagram of the Gross-Neveu model in 1+1 dimensions is studied using Ginzburg-Landau expansion. It predicts several features of the exact phase diagram correctly even at low orders. It is shown that increasing the order of the expansion improves the accuracy of the crystal phase except for very small temperatures, where the expansion completely fails regardless of the order of the expansion. The source of this behaviour seems to be related to the Silver-Blaze phenomenon.}




\maketitle

%%=============================================================%%
%%=============================================================%%
%%=============================================================%%
\section{Introduction}\label{sec1}
%%=============================================================%%
%%=============================================================%%
%%=============================================================%%
The Gross-Neveu model is a quantum field theory of $N_f$ flavours of massless fermions with a four-fermion interaction term \cite{grossNeveu}
\begin{equation} \label{eq:GNLag}
	\mathcal{L} = \bar{\psi_a} i  \slashed{\partial} \psi_a + \frac{g^2 }{2}  (\bar{\psi_a}\psi_a)^2, \quad a = 1,2,.., N_f.
\end{equation}
Despite its simplicity the 1+1 dimension version of the model (henceforth GN$_2$), possesses several interesting properties in the 't Hooft limit ($N_f \rightarrow \infty$ with $g^2 N_f$ held constant). The model shares several important features with Quantum Chromodynamics, such as asymptotic freedom, renormalizability and dynamical breakdown of the discrete chiral symmetry.  This last property is responsible for giving a dynamical mass to the otherwise massless fermions.

The Lagrangian of the model can be rewritten in terms of an auxiliary field $\phi$ by means of a Hubbard-Stratanovich transformation as
\begin{equation} \label{eq:GNLagHubStrat}
	\mathcal{L} = \bar{\psi} \left(i  \slashed{\partial} - \phi \right) \psi - \frac{1}{2g^2} \phi^2.
\end{equation}
where the sum over the flavour index $a$ has been suppressed. The auxiliary field $\phi$ breaks the discrete chiral symmetry. It satisfies a consistency condition in the 't Hooft limit
\begin{equation}
	\phi = - g^2 \langle \bar{\psi} \psi \rangle
\end{equation}
which suggests a physical interpretation of $\phi$ as a bosonic condensate. The fermionic fields can be integrated out, as the Lagrangian is quadratic in them, leaving behind an effective action
\begin{align}
	S_\text{eff} =  - i \ln \det \left(  i  \slashed{\partial} - \phi\right) -\frac{1}{2g^2 N_f} \int d^2x \,\phi^2.
\end{align}

In the 't Hooft limit the model can be studied by solving a \emph{gap equation} 
\begin{align} \label{eq:gap}	
	&\frac{\delta S_\text{eff}}{\delta \phi} = 0 \nonumber \\
	&\implies \frac{1}{g^2 N_f} \phi + i \frac{\delta}{\delta \phi} \ln \det \left(  i  \slashed{\partial} - \phi\right) =0.
\end{align}

For equilibrium thermodynamics, only static condenstates are considered, $\partial \phi / \partial t  = 0$. Early work on the phase diagram of the GN$_2$ model assumed a spatially homogeneous condensate \cite{wolff}. This assumption simplifies what is otherwise a much harder problem to solve, but misses certain important features of the correct phase diagram such as the correct order of phase transitions as well as the existence of a crystalline phase \cite{thiesOriginal,gokceOrig}. The correct phase diagram was obtained via a relativistic Hartree-Fock approach in \cite{thiesOriginal} and by directly solving the gap equation in \cite{gokceOrig}. These two approaches are equivalent. Additionally, see \cite{thies-review} for a review.

The gap equation has the solution \cite{thiesOriginal}
\begin{equation} \label{eq:condensateOriginal}
	\phi(x) = m \tilde{\nu} \,\dfrac{\text{sn} (mx; \tilde{\nu})\, \text{cn} (mx; \tilde{\nu})}{\text{dn} (mx; \tilde{\nu})}
\end{equation}
where sn, cn and dn are the standard Jacobi elliptic functions. An equivalent expression is \cite{gokceOrig}
\begin{equation} \label{eq:condensateSimpleLambda}
	\phi(x) = \lambda \sqrt{\nu} \, \text{sn} \left( \lambda x;\nu \right).
\end{equation}
The two expressions are connected by a Landen transformation \cite{DLMFLanden} and the mapping
\begin{equation}
	\lambda = \dfrac{2m}{1+\sqrt{\nu}}, \qquad 	\tilde{\nu} = \dfrac{4 \sqrt{\nu}}{\left( 1+\sqrt{\nu} \right)^2}.
\end{equation}
The physical interpretation of the parameters is simple: the scale parameter $\lambda$ sets the size or amplitude of the condensate and the elliptic parameter $\nu$ governs the shape of the condensate.

To find the thermodynamic phase at a given temperature and chemical potential, one needs to minimize the grand canonical potential, denoted $\Psi$ or GCP henceforth, with respect to the self-consistent condensate $\phi(x)$ which serves as the (inhomogeneous) order parameter for this model.

Due to the simplicity of the solution \eqref{eq:condensateSimpleLambda}, the task of minimization of GCP with respect to $\phi(x)$ reduces to a much easier one of minimization with respect to only two parameters $\lambda$ and $\nu$.
\begin{equation}
	\frac{\delta}{\delta \phi} \Psi[\phi(x)] = 0 \xRightarrow{\scriptstyle{\phi = \lambda \sqrt{\nu}\, \text{sn}(\lambda x,\nu)}} \
	\begin{array}{l}
		\partial_\lambda \Psi(\lambda,\nu) = 0 \\ 
		\partial_\nu \Psi(\lambda,\nu) = 0
	\end{array} .
\end{equation}
While it is intuitively obvious that minimization with respect to either of the pair of variables, $(\lambda, \nu)$ or $(m, \tilde{\nu})$, will result in the same phase diagram, one can evaluate the Jacobian explicitly and verify that this is indeed the case.

\begin{figure}
	\centering
	\includegraphics[width=0.48\textwidth]{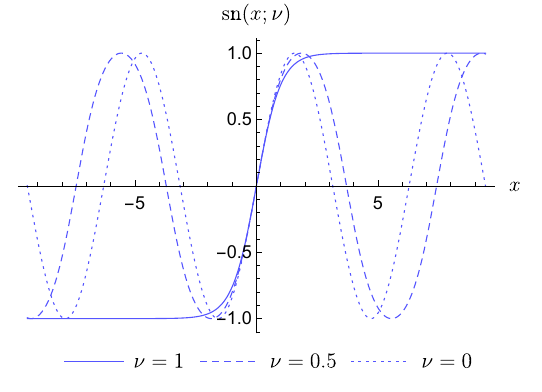}
	\caption{\small{The Jacobi elliptic sine function sn for several values of the elliptic parameter $\nu$. The function is a pure sine at $\nu=0$. The period increases with $\nu$ and eventually at $\nu=1$ the function loses its periodicity and converts into a kink ($\tanh x$).}} \label{fig:jacobi}
\end{figure}

The minimizing values of these parameters govern the phase of the GN$_2$ baryonic matter at any given point in the $\mu$-$T$ plane. This association of phase with the values of $\lambda$ and $\nu$ (or $m$ and $\tilde{\nu}$) is simple and follows from the properties of the sn function (see Figure \ref{fig:jacobi}):
\begin{itemize}\setlength\itemsep{-0.5em}
	\item $\lambda = 0$ or $\nu=0$ : chirally and translationally symmetric (massless homogeneous phase) \\ 
	\item $\lambda >0$, $\nu=1$	: chiral symmetry broken, translationally symmetric (massive homogeneous phase) \\
	\item $\lambda >0$, $0 < \nu < 1$ : both chiral and translational symmetry broken (massive crytalline phase).
\end{itemize}
%\begin{center}
%	\begin{tabular}{ |l|l| } 
%		\hline
%		$\lambda = 0$ & chirally translationally symmetric \\
%		 or $\nu=0$ & (massless homogeneous phase) \\ 
%		\hline
%		$\lambda >0$, & chiral symmetry broken, translationally symmetric \\
%		$\nu=1$		& (massive homogeneous phase) \\ 
%		\hline
%		$\lambda >0$,& both chiral and translational symmetry broken \\
%		$0 < \nu < 1$ & (massive crytalline phase) \\ 
%		\hline
%	\end{tabular}
%\end{center}

The GCP is \cite{thies2004}
\begin{equation}\begin{aligned}
		\Psi &= \dfrac{m^2 Z \ln \Lambda}{2\pi} -\dfrac{1}{\pi \beta}  \int_\text{band} dE \left|\dfrac{dq}{dE} \right| \\
		& \quad \times \ln \bigg[ \left(1 + e^{-\beta (E - \mu)} \right) \left(1 + e^{\beta (E + \mu)} \right) \bigg] 
\end{aligned} \end{equation}
where the energy band is $[0,m X] \cup [m,E_\Lambda]$, $\Lambda$ is the cut-off, $q(E)$ is the band momentum with the derivative 
\begin{equation} \label{eq:dispersionDerivative}
	\dfrac{dq}{dE} = \pm \dfrac{ E^2 - m^2Y^2}{\sqrt{ (E^2-m^2X^2) (E^2-m^2)}}.
\end{equation}
The various symbols are defined as
\begin{equation} \begin{aligned}
		&E_\Lambda = \dfrac{\Lambda}{2} + \dfrac{ m^2 Z}{\Lambda} + \dots,\\
		&X = \dfrac{1-\sqrt{\nu}}{1+\sqrt{\nu}} = \sqrt{1- \tilde{\nu}}, \\
		&Y =  \dfrac{\sqrt{1 + \nu - 2 \left(1 - \mathbf{E}/\mathbf{K}\right)}}{1+\sqrt{\nu}} = \sqrt{\dfrac{\tilde{\mathbf{E}}}{\tilde{\mathbf{K}}}},\\
		&Z = \dfrac{4 (1 - \mathbf{E}/\mathbf{K})}{(1+\sqrt{\nu})^2}\ = 2 \left( 1 - \dfrac{\tilde{\nu}}{2} - \dfrac{\tilde{\mathbf{E}}}{\tilde{\mathbf{K}}} \right).
\end{aligned}\end{equation}
The shorthands $\mathbf{K}$ and $\tilde{\mathbf{K}}$ stand for the complete elliptic integrals of first kind $\mathbf{K}(\nu)$ and $\mathbf{K}(\tilde{\nu})$ respectively, defined by $\mathbf{K}(\nu) = \int_0^{\pi/2} d\theta / \sqrt{1- \nu \sin^2 \theta}$. The shorthands $\mathbf{E}$ and $\tilde{\mathbf{E}}$ are similarly defined in terms of the complete elliptic integrals of second kind, $\mathbf{E}(\nu) = \int_0^{\pi/2} \sqrt{1- \nu \sin^2 \theta} \, d\theta$.

The GCP is clearly UV-divergent and the renormalized GCP, obtained after dropping irrelevant divergent terms, is
\begin{align} \label{eq:renormalizedGCP}
		\Psi_\text{ren} &= \dfrac{m^2 }{4\pi} \left( (1-X)^2 - 2Z \right)   + \dfrac{m^2 Z}{2\pi} \ln \dfrac{2m }{1+\sqrt{\nu}} \nonumber \\
		& -\dfrac{1}{\pi\beta}  \int_\text{band} dE \left|\dfrac{dq}{dE} \right| \ln \Big[ \left(1 + e^{-\beta (E- \mu)} \right)  \nonumber \\
		& \qquad\qquad\qquad \times \left( 1 + e^{-\beta (E + \mu)}\right) \Big] .
\end{align}

Another way to study the GN$_2$ model is through a Ginzburg-Landau expansion of the GCP \cite{gokceOrig,bdt}. Typically Ginzburg-Landau expansions are phenomenological anzatzes based on symmetries of the system. For example, \cite{bogdanov,robler} predict skyrmions in magnetic crystals from a Ginzburg-Landau expansion constructed from the symmetries of these crystals. Ginzburg-Landau expansions can also be generated for higher dimensional Gross-Neveu models, where explicit exact solutions do not exist \cite{nickel}. Such expansions take the form of derivative expansions, with the generic form
\begin{equation}
	\Psi_\text{GL} = \sum_n c_n(T,\mu) \, F_n [\phi, \vec{\nabla} \phi, \dots ].
\end{equation}
GN$_2$ is a rare case where one can generate a Ginzburg-Landau expansion directly from the underlying microscopic theory; each term in the expansion can be calculated exactly. Thus, GN$_2$ can be viewed as a test model where phase diagrams obtained from Ginzburg-Landau expansions can be studied at various orders and compared against the exact phase diagram. The phase diagram generated from even a low order expansion (keeping first two non-trivial terms only) exhibits several crucial features of the exact phase diagram such as the existence of a crystalline phase and a tri-critical point, along with expressions for some of these features that match perfectly with the exact phase diagram \cite{bdt}. The following section discusses the GN$_2$ Ginzburg-Landau expansion in more detail.

%%=============================================================%%
%%=============================================================%%
\section{Ginzburg-Landau expansion for GN$_2$} \label{sec2}
%%=============================================================%%
%%=============================================================%%

The Ginzburg-Landau expansion for the renormalized GCP of the GN$_2$ model for any arbitrary static condensate $\phi(x)$ is given by \cite{thies,bdt} 
\begin{align} \label{eq:OPExpGCP}
		\Psi_\text{GL} &= \alpha_0(T,\mu) + \alpha_2(T,\mu) \langle\phi^2\rangle + \alpha_4(T,\mu)  \langle\phi^4 + \phi'^2\rangle \nonumber \\
		 & +  \alpha_6(T,\mu)  \langle\phi^6  + 5 \phi^2 \phi'^2  + \frac{1}{2} \phi''^2 \rangle+ \dots
\end{align}
where prime indicates spatial derivative and $\langle \dots \rangle$ indicates average over one period of $\phi(x)$. The spatial derivatives  in the expansion are crucial, as they are responsible for the spatially inhomogeneous phase that is known to exist in the exact GN$_2$ phase diagram.

Minimization of the Ginzburg-Landau GCP at higher orders is a formidable problem. A remarkable property of the GN$_2$ model comes to rescue: the condensate \eqref{eq:condensateSimpleLambda} solves the Ginzburg-Landau equation $$\frac{\delta \Psi_\text{GL}}{\delta \phi(x)} = 0$$ at every order of the expansion.\footnote{This is a result of certain integrability properties of the GN$_2$ gap equation: the set of Ginzburg-Landau equations, one equation at each order, forms a heirarchy of differential equations known as the modified Korteweg-de Vries hierarchy \cite{gokceOrig,bdt, belokolos}.} This means that instead of solving a different differential equation at every order, one can simply evaluate the Ginzburg-Landau GCP \eqref{eq:OPExpGCP} on the condensate \eqref{eq:condensateSimpleLambda} to obtain an expansion that depends on only two parameters $\lambda$ and $\nu$. The expansion is
\begin{equation} \label{eq:GLExpAlln}
	\Psi^{(N)} = \frac{1}{2\pi}\sum_{n=0}^{N} \lambda^{2n} \alpha_{2n}(T,\mu) f_{2n}(\nu).
\end{equation}
This expansion can now be minimized with respect to $\lambda$ and $\nu$, a much simpler task. Note that the dependence on $\lambda$ and $\nu$ has cleanly separated. Also note that the expansion is a small $\lambda$ expansion, so there is a natural expectation that the resulting phase diagram is accurate close to the massless ($\lambda=0$) phase boundary, and that the accuracy may fall off further from the massless phase boundary.

The $\nu$-dependence from the spatial averages is collected into the functions $f_{2n}(\nu)$ which are normalized as $f_{2n}(1) = 1$. The appearance of these functions is a result of terms with derivatives of $\phi$ in the expansion \eqref{eq:OPExpGCP}. On a side note, one can recover the Ginzburg-Landau expansion corresponding to a homogeneous condensate by taking the limit $\nu \rightarrow 1$ which sets all $f_{2n}$ to unity. These functions increase monotonically with $\nu$ and have the generic form $$f_{2n}(\nu) = p^{(2n-2)}(\nu) + q^{(2n-2)}(\nu) \dfrac{\mathbf{E}}{\mathbf{K}}$$ where $p^{(n)}(\nu)$ and $q^{(n)}(\nu)$ are polynomials of order $n$. 

The expansion \eqref{eq:GLExpAlln} can also be generated directly from the GCP \eqref{eq:renormalizedGCP} by expanding $dq/dE$ in powers of $\lambda/E$ \cite{thies2004}. This process also produces a recursion relation for generating $f_{2n}(\nu)$. See Appx. \ref{secA1} for the recursion relation as well as explicit expressions for small $n$.

The temperature and chemical potential dependence in the expansion is through the functions $\alpha_{2n}(T,\mu)$, which are given by
\begin{align} 
		&\alpha_0 = -\frac{\pi^2 T^2}{3} - \mu^2 \nonumber\\
		&\alpha_{2} =  \text{Re}\, \psi^{(0)}\left(\frac{1}{2} + \frac{i\mu}{2\pi T} \right) +  \ln \,(4\pi T)  \\
		&\alpha_{2n} = \frac{(-1)^{n-1} }{n!(n-1)! (4 \pi T )^{2n-2}} \nonumber \\
		& \qquad  \times \text{Re} \, \psi^{(2n-2)}\left(\frac{1}{2} + \frac{i\mu}{2\pi T} \right), \qquad n \geq 2, \nonumber
\end{align}
where $\psi^{(n)}$ is the polygamma function of order $n$. Polygamma functions with identical $\mu$ and $T$ dependence appear naturally in expansions of thermodynamics quantities of the free Fermi gas \cite{blinnikov}. It is through the log and quadratic temperature dependence in $\alpha_0(T,\mu)$ and $\alpha_2(T,\mu)$ that the four-fermion interaction of the GN$_2$ model manifests itself.

%%=============================================================%%
%%=============================================================%%
\section{Phase diagrams from low order Ginzburg-Landau expansions}\label{sec3}
%%=============================================================%%
%%=============================================================%%

The $N$-th order phase diagram is computed by minimizing the $N$-th order GCP \eqref{eq:GLExpAlln}, that is, by simultaneously solving
\begin{align} \label{eq:minCond}
	\partial_\nu \Psi^{(N)} = 0 \qquad \text{and } \qquad \partial_\lambda \Psi^{(N)}= 0.
\end{align}
The GCP at leading order is
\begin{equation}
	\Psi^{(1)} = \alpha_0 + \lambda^2 \alpha_2 f_2(\nu).
\end{equation}
This obviously has only one minimum, at $\lambda = 0$, but only in the region $\alpha_2(T,\mu) >0$, and none in $\alpha_2(T,\mu) < 0$. Thus, at leading order, the phase diagram is trivial and severely incomplete. It has a single phase, the massless homogeneous phase, and no phase transitions.

%%=============================================================%%
\subsection{Second order phase diagram} %%=============================================================%%
The first signs of any phase transitions are seen at second order, at which order the GCP is 
\begin{equation} \label{eq:GCPSecondOrder}
	\Psi^{(2)} = \alpha_0 + \lambda^2 \alpha_2 f_2(\nu) + \lambda^4 \alpha_4 f_4(\nu).
\end{equation}

\begin{figure}
	\centering
	\includegraphics[width=0.47\textwidth]{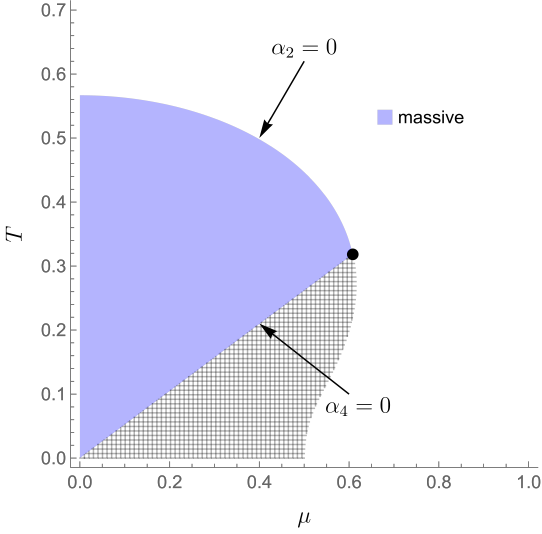}
	\caption{\textbf{Second order phase diagram}: dot - tricritical point, white - massless phase, mesh -  analysis is invalid as $\Psi^{(2)}$ is unbounded from below. The tri-critical point is actually not a feature of the phase diagram at second order, but is included here for comparison and visualization purposes.} \label{fig:SecondOrderPhaseDiagram}
\end{figure}

A single minimum, $\lambda = 0 $, exists in the region $\alpha_2(T,\mu) > 0$. This is the massless homogeneous phase. A global (but not local) minimum 
\begin{equation}
	\lambda = \sqrt{- \frac{\alpha_2}{2 \alpha_4}}, \qquad \nu = 1
\end{equation}
exists in the region $\alpha_2(T,\mu) <0$,  $\alpha_4(T,\mu) >0$. This is the massive homogeneous phase. No other phases exist at this order. The phase boundary between the two phases is given by the curve $\alpha_2(T,\mu) = 0$ constrained to the region $\alpha_4(T,\mu) > 0$. This phase boundary matches perfectly with the exact phase diagram, and does not change as the order of the expansion is increased. See Fig. \ref{fig:SecondOrderPhaseDiagram} for the phase diagram. 

%%=============================================================%%
\subsection{Third order phase diagram}
\begin{figure} [t]
	\centering
	\includegraphics[width=0.48\textwidth]{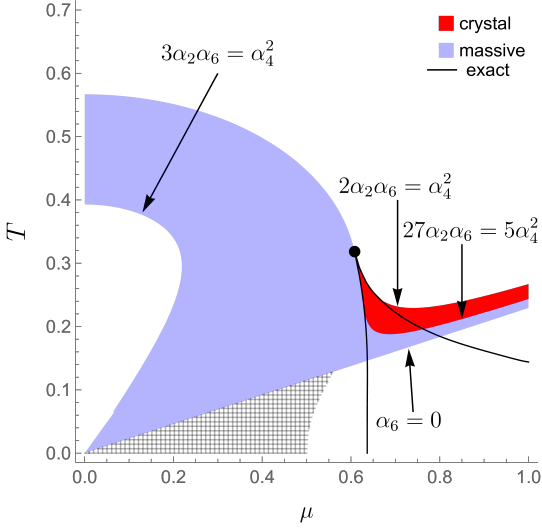}
	\caption{\textbf{Third order phase diagram}: white - massless phase, dot - tricritical point, mesh - analysis is invalid because $\Psi^{(3)}$ is unbounded from below. Solid line shows the boundaries of the exact numerically computed crystal phase. See Fig. \ref{fig:ThirdOrderPhaseDiagramZoomed} for a zoom in on the crystal phase.} \label{fig:ThirdOrderPhaseDiagram}
\end{figure}

%%=============================================================%%
The expansion at this order is
\begin{equation} \label{eq:GCPO6}
	\Psi^{(3)} = \alpha_0 + \lambda^2 \alpha_2 f_2(\nu) + \lambda^4 \alpha_4 f_4(\nu) + \lambda^6 \alpha_6 f_6(\nu).
\end{equation}
In addition to the massless and massive homogeneous phases, a third minimum is found at this order. The minimizing parameters $\lambda$ and $\nu$ are given implicitly by
\begin{equation} \label{eq:O6MinValLambda}
	\lambda  = \sqrt{- \left( \frac{3f_6f_2' - f_2f_6'}{3f_6f_4' - 2f_4f_6'} \right) \frac{\alpha_2}{\alpha_4}}, 
\end{equation}
\begin{equation} \label{eq:O6MinValNu}
	\dfrac{\alpha_2 \alpha_6}{\alpha_4^2} = \frac{\left(2 f_4 f_2' - f_2 f_4' \right)\left(3 f_6 f_4' - 2 f_4 f_6'\right) }{\left( f_2 f_6' - 3 f_6 f_2'\right)^2},
\end{equation}
and restricted to the region $\alpha_2(T,\mu) > 0$, $\alpha_4(T,\mu) < 0$. This is the crystal phase, with broken translation invariance due to a non-zero scale parameter $\lambda$ and elliptic parameter $\nu$ between 0 and 1. See Fig. \ref{fig:ThirdOrderPhaseDiagram} for the phase diagram. 

\eqref{eq:O6MinValNu} is an implicit equation for the minimizing elliptic parameter $\nu = \nu(T,\mu)$, which can be thought of as describing an infinite family of curves (in the $\mu$-$T$ plane) inside the crystal phase, each curve associated with a particular value of $\nu$. These curves span the entire crystal phase. The boundary between massless and crystal phases ($\nu \rightarrow 0$ in \eqref{eq:O6MinValNu}) is
\begin{equation}
	\alpha_2(T,\mu) \alpha_6(T,\mu) = \frac{1}{2} \left(\alpha_4(T,\mu)\right)^2,
\end{equation}
while that between crystal and massive homogeneous phases ($\nu \rightarrow 1$ in \eqref{eq:O6MinValNu}) is
\begin{equation}
	\alpha_2(T,\mu) \alpha_6(T,\mu) = \frac{5}{27} \left(\alpha_4(T,\mu)\right)^2.
\end{equation}
The phase boundary betweeen massless and massive homogeneous phases is same as before: $\alpha_2(T,\mu) = 0,\, \alpha_4(T,\mu) > 0$. All three phases meet at a tricritical point given by the intersection of the massive-massless phase boundary and the crystal phase:
\begin{equation} \label{eq:tricrit}
	\alpha_2(T,\mu) = 0 = \alpha_4(T,\mu).
\end{equation}
This result agrees perfectly with that from the exact solution. The sliver of crystal phase near the tricritical point agrees very well with the exact crystal phase, but then as $\mu$ (or $T$) increases it diverges away, running asymptotic to the lower of the two lines on which $\alpha_6(T,\mu)$ vanishes.  

\begin{figure}
	\includegraphics[width=.48\textwidth]{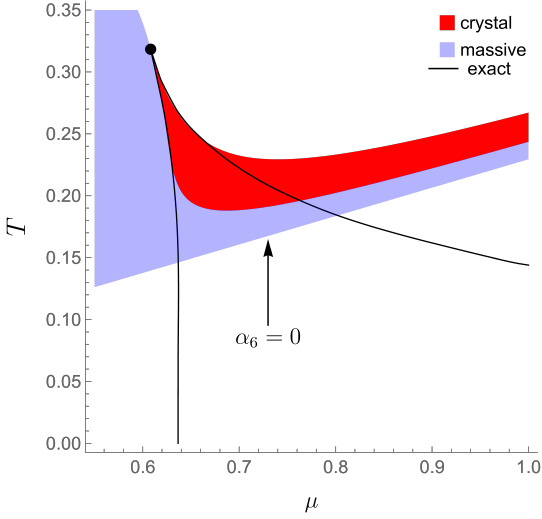}
	\caption{\textbf{Third order phase diagrams zoomed in on the crystal region}: dot - tricritical point, white - massless phase,  solid line - exact boundary of the crystal phase}\label{fig:ThirdOrderPhaseDiagramZoomed} 
\end{figure}

\begin{figure} [!t]
	\includegraphics[width=0.48\textwidth]{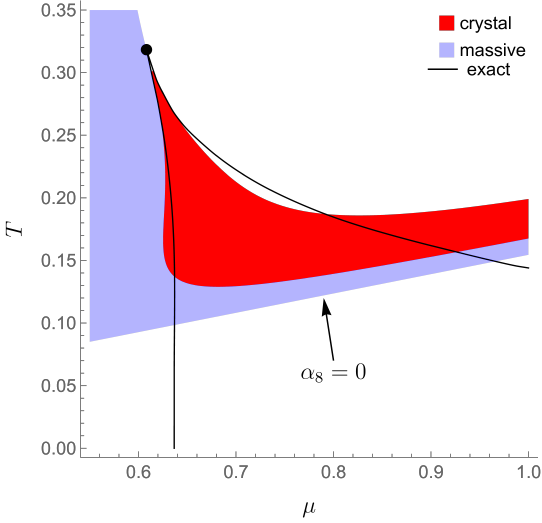} \includegraphics[width=0.48\textwidth]{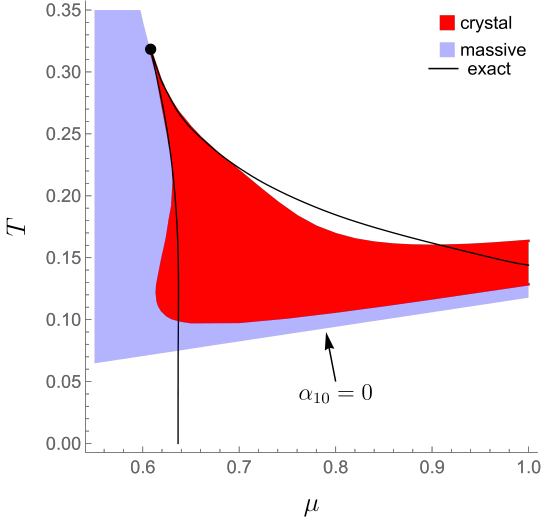}
	\caption{\textbf{Fourth (left) and fifth (bottom) order phase diagrams zoomed in on the crystal region}: dot - tricritical point, white - massless phase,  solid line - exact boundary of the crystal phase. Increasing the order clearly improves the approximation. Note that the crystal phase in each case is asymptotic to a line on which the highest order coefficient vanishes.}\label{fig:FourthFifthOrderPhaseDiagramZoomed}
\end{figure}

A final feature to note here is the existence of a spurious massless phase inside the massive homogeneous phase, whose boundary is given by $3 \alpha_2(T,\mu)\alpha_6(T,\mu) = \alpha_4(T,\mu)^2$. This is in complete disagreement with the exact phase diagram. In fact, at higher orders, numerous fragmented spurious phases appear in the region which should have only the massive homogeneous phase. For this reason, analysis of higher order phase diagrams will be restricted to the crystal phase, which is proximate to the massless phase, and does not suffer from this issue.

%%=============================================================%%
\subsection{Fourth and fifth order phase diagrams}
%%=============================================================%%
Going to the fourth order improves the resemblance of the crystal phase to the exact crystal phase. The phase becomes broader, dips closer to the $\mu$-axis and agrees with the exact crystal phase further away from the tricritical point than the third order crystal phase does.

At this order, the equation of the crystal phase is too cumbersome to be given here in full detail. The equations for the phase boundaries are simpler. The massless-crystal phase boundary is
\begin{equation} \begin{aligned}
		\dfrac{54}{5} \alpha^2_2 \alpha_8   + \alpha_4 \left(4 \alpha_4^2 - 9 \alpha_2 \alpha_6 \right) = \dfrac{1}{2} \left(4 \alpha_4^2 - 6 \alpha_2 \alpha_6 \right)^{3/2}
\end{aligned}\end{equation}
and the crystal-massive phase boundary is 
\begin{align}
		&\dfrac{7290}{7} \alpha^2_2 \alpha_8 + \alpha_4 \left(40 \alpha_4^2 - 243 \alpha_2 \alpha_6 \right) \nonumber \\
		& \qquad = \dfrac{1}{25} \left(100 \alpha_4^2 - 405 \alpha_2 \alpha_6\right)^{3/2}
\end{align}
both constrained to the region $\alpha_2(T,\mu) > 0$, $\alpha_4(T,\mu) < 0$. 

The fifth order phase diagram has an even larger crystal phase, and is accurate to the true crystal phase even further from the tri-critical point. See Figure \ref{fig:FourthFifthOrderPhaseDiagramZoomed} for crystal phases at both fourth and fifth orders. It is evident that increasing the order of the Ginzburg-Landau expansion improves the accuracy of the crystal phase. The equations of the crystal region as well as the phase boundaries are very cumbersome at this order and are left out.

A final point to note is that the crystal region in every case so far has been asymptotic to a straight line in the $\mu$-$T$ plane, and never touches the $\mu$-axis. The pattern is apparently this: at $N$-th order the crystal phase is asymptotic a straight line satisfying $\alpha_{2N}(T,\mu) = 0 $. In the next section it will shown that this is true all $N \geq 3$.

%%=============================================================%%
%%=============================================================%%
\section{Higher order phase diagrams}\label{sec4}
%%=============================================================%%
%%=============================================================%%
Before proving the result proposed at the end of the previous section, it is instructive to work out the asymptotic behavior of the third order crystal phase in detail, as the essentials features are already present there.

%%=============================================================%%
\subsection{A simple exercise on the third order crystal phase}
%%=============================================================%%
With the shorthand
\begin{align}
	& R_n(z) \equiv \text{Re } \psi^{(n)}\left(\frac{1}{2} + \frac{iz}{2 \pi} \right), \nonumber \\
	& I_n(z) \equiv \text{Im } \psi^{(n)}\left(\frac{1}{2} + \frac{iz}{2 \pi} \right),
\end{align}
the equation of the crystal phase at third order \eqref{eq:O6MinValNu} can be rewritten as
\begin{equation} \label{eq:O6explicit}
		\left(R_0 \left(\frac{\mu}{T}\right) + \ln (4\pi T) \right) \dfrac{R_4 \left(\frac{\mu}{ T}\right)}{3R_2^2 \left(\frac{\mu}{T}\right)} =  F(\nu) 
\end{equation}
where $F(\nu)$ is the expression on the right hand side of \eqref{eq:O6MinValNu}. A straight line (denoted $L_3$) of the form $T = \mu/c_3$ satisfies $\alpha_6(T,\mu) = 0$ (equivalent to $R_4(\mu/T) = 0$). In the vicinity of $L_3$, $T$ and $\mu$ are trivially related as 
\begin{equation}
	c_3 T = \mu ( 1+ \epsilon_3(\mu))
\end{equation}
where $\epsilon_3(\mu) \ll 1$ is a measure of the vertical distance of the crystal phase from $L_3$.\footnote{The awkward subscripts in $L_3$, $c_3$, $\epsilon_3$ ensure consistency with the notation used later.} Substituting this in \eqref{eq:O6explicit}, expanding about $\epsilon_3 = 0$ to leading order, and solving for $\epsilon_3$ gives 
\begin{equation} \label{eq:epsilonForOrderThreeFull}
	\epsilon_3(\mu) = \dfrac{3 \pi F(\nu) \left(R_2 (c_3) \right)^2  }{c_3 \left( R_0 (c_3) + \ln \frac{4\pi\mu}{_3c} \right) I_5(c_3)}
\end{equation}
which for large $\mu$ scales as
\begin{equation} \label{eq:epsilonScalingOrderThree}
	\epsilon_3(\mu) \propto \dfrac{1}{\ln \mu}.
\end{equation}
This is consistent with the initial assumption that \eqref{eq:O6explicit} can be expanded about $\epsilon_3(\mu) = 0$. This scaling implies that the crystal phase is asymptotic to $L_3$. The equation for the minimizing $\lambda$ \eqref{eq:O6MinValLambda} implies that for large $\mu$ in the vicinity of $L_3$
\begin{equation} \label{eq:O6lambdascaling}
	\dfrac{\lambda}{T} \propto \dfrac{1}{\sqrt{\epsilon_3(\mu)}} 
\end{equation}
which implies $\lambda/T$ scales as $\sqrt{\ln \mu}$.

Now, there are actually two lines which satisfy $\alpha_6(T,\mu) = 0$, not one. Their slopes are $1/c_3 \approx 0.976, 0.229$. The label $L_3$ will now be used to denote the line with smaller slope, while $L_3'$ will denote the steeper line. Using these numerical values in \eqref{eq:epsilonForOrderThreeFull} gives the deviation from each line, respectively, as
\begin{align}\label{eq:epsilonScalingOrderThreeNumeric}
	&L_3' : \epsilon_3(\mu) \approx -\dfrac{0.133}{0.747 + \ln \mu} \nonumber \\
	&L_3 :  \epsilon_3(\mu) \approx \dfrac{0.096}{0.583 + \ln \mu}.
\end{align}
The first of these is negative for large $\mu$ and is rejected to avoid an imaginary $\lambda$ because of \eqref{eq:O6lambdascaling}. The second expression in \eqref{eq:epsilonScalingOrderThreeNumeric} is valid, showing that the crystal phase is asymptotic to $L_3$ but not $L_3'$, that is, to the lower of the lines satisfying $\alpha_6(T,\mu) = 0$.

On $L_3$ the GCP reduces to the second order GCP, $\Psi^{(2)}$. Therefore, based on the minimization at second order, only the massless homogeneous phase can be present on $L_3$. Combining all of these statements together leads to the result that the third order crystal phase always stays above the lowest of the lines satisfying $\alpha_6(T,\mu) = 0$ and thus can never touch the $\mu$-axis.

%%=============================================================%%
\subsection{Expansions of arbitrary orders ($N \geq 3$)}
%%=============================================================%%
The qualitative behaviour of the crystal phase remains the same if the order is increased beyond $N=3$, including the fact that it never touches the $\mu$-axis. The proof follows the same general steps in the example above.

\begin{figure} [!t]
	\centering				
	\includegraphics[width=0.48\textwidth]{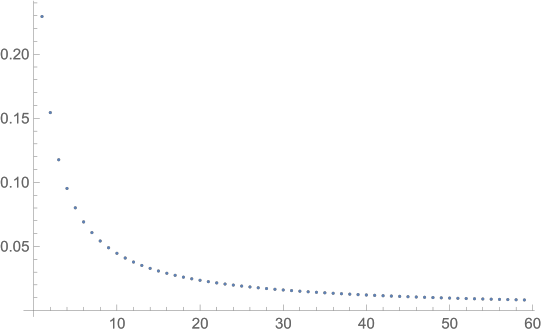} 
	\caption{Smallest root of the function $R_{2N}(1/z) = \text{Re } \psi^{(2N)}\left(\dfrac{1}{2} + \dfrac{i}{2 \pi z} \right)$ as a function of $N (\geq 3)$}
	\label{fig:roots}
\end{figure}

The equation $\alpha_{2N}(T,\mu) = 0$ is satisfied by $N-1$ straight lines in the $\mu$-$T$ plane. Suppose $1/c_N$ is the slope of the lowest of these lines, denoted $L_N$. The other lines are denoted as a whole by $L_N'$. Then $c_N$ is also the smallest of the $N-1$ roots of $R_{2N-2}(1/z)$. The (scaled) deviation of the crystal phase from $L_N$ is
\begin{equation}\label{eq:NthOrderDeviation} 
	\epsilon_N(\mu) = \dfrac{c_N \, T - \mu}{\mu}.
\end{equation}
It can be shown that the $N$-th order minimization conditions imply, for every $\nu$ and large $\mu$
\begin{equation} 	\label{eq:largeOrderScaling} 
	\dfrac{\lambda}{T} \propto \dfrac{1}{\sqrt{\epsilon_N(\mu)}} , \quad
	\epsilon_N(\mu) \propto \dfrac{1}{N}\left(\frac{1}{\ln \mu}\right)^{\frac{1}{N-2}}
\end{equation}
These scalings imply that the $N$-th order crystal phase is asymptotic to the lowest line satisfying $\alpha_{2N}(T,\mu) = 0$. See Appx. \ref{secA2} for details. For completeness we provide the full expression, obtained from \eqref{eq:MinCondAnyNWithoutAlpha2N}, of the distance between the lower edge of the crystal phase ($\nu=1$) and $L_N$:
\begin{equation}\label{eq:deviationfullExpression}\begin{aligned} 
		& \epsilon_N(\mu) = \dfrac{2 \pi}{c_N} \left|\frac{R_{2N-4}(c_N)}{I_{2N-1}(c_N)} \right|  \dfrac{(N-1)(N-2)(2N-1)}{2N-3} \\
		& \times \left(\dfrac{R_{2N-4}(c_N)}{(N-1)! (N-2)! (2N-3) \left(\ln \frac{4\pi \mu}{c_N} + R_0(c_N) \right)} \right)^\frac{1}{N-2}
\end{aligned}\end{equation}
This asymptotic property of the crystal phase is, as it is, not very interesting. Using induction, with similar logic as discussed in the example above, it can be easily proved that there is no crystal phase on $L_N$ - only a massless phase. A third fact is that the slope of the asymptote $L_N$ decreases with $N$ (see Figure \ref{fig:roots}). Combining all three results implies that the $N$-th order crystal phase dips lower and lower as $N$ increases, just as expected, but it can never touch the $\mu$-axis for any $N$. The conclusion is that the Ginzburg-Landau expansion completely fails to predict the zero temperature phase diagram, regardless of the order of the expansion, as it predicts a massless homogeneous phase at zero temperature.

%%=============================================================%%
%%=============================================================%%
\section{Small temperature behaviour and failure of the Ginzburg-Landau approach}
%%=============================================================%%
%%=============================================================%%
As the Ginzburg-Landau approach fails to provide a zero temperature phase diagram even remotely similar to the correct phase diagram, it is worth looking at the GCP at small temperatures.

The exact zero temperature GCP can be obtained by taking the zero temperature limit of \eqref{eq:renormalizedGCP}, and is a piecewise continuous function. It consists of three separate expressions, each corresponding to whether the chemical potential lies in the inner band $0 \leq \mu \leq m X$, the gap $ m X < \mu < m$ or the outer band $\mu \geq m$, or equivalently whether $m$ is large, medium or small, respectively. The full expression is

\begin{strip}
\rule{0.5\textwidth}{.2mm} \\
\begin{align} \label{eq:zeroTGCP}
	&\Psi_{T=0} =\nonumber \\
	& \begin{cases}
		\begin{aligned}			
			&	\dfrac{m \mu}{\pi}\, G\left(\dfrac{\mu}{m X} \right) - \dfrac{1}{2 \pi} \sqrt{m^2 - \mu^2} \sqrt{m^2 X^2 - \mu^2} + \dfrac{m^2 H}{4\pi} 
			+ \dfrac{m^2 Z}{2\pi} \ln \left( \sqrt{m^2 - \mu^2} + \sqrt{m^2 X^2- \mu^2} \right), \quad \text{if } m \geq \dfrac{\mu}{X} \\[1ex]
			&-\dfrac{m \mu}{2 \mathbf{K}(\tilde{\nu})} + \dfrac{m^2 H}{4\pi} + \dfrac{m^2 Z}{4\pi} \ln (m^2 \tilde{\nu}), \quad  \text{if } \dfrac{\mu}{X} \leq m \leq \mu \\[1ex]
			& \dfrac{m \mu}{\pi}\, G\left(\dfrac{m}{\mu} \right) - \dfrac{1}{2\pi}\sqrt{\mu^2-m^2}\sqrt{\mu^2-m^2 X^2} + \dfrac{m^2 H}{4\pi} + \dfrac{m^2 Z}{2\pi} \ln \left(\sqrt{\mu^2 - m^2X^2} + \sqrt{\mu^2 - m^2} \right), \quad \text{if } m \leq \mu
		\end{aligned}
	\end{cases}
\end{align}
where
\begin{align}
	G(z) = (1- Y^2)\textbf{F} \left( \arcsin z,X^2 \right) - \textbf{E} \left(\arcsin z,X^2 \right), \qquad H = 1+X^2 - 2Z,
\end{align}
and $\mathbf{F}$ and $\mathbf{E}$ are the incomplete elliptic integrals of first and second kind, respectively, defined by
\begin{align}
	\mathbf{F}(\eta,k) = \int_0^{\eta} \dfrac{d\theta}{\sqrt{1- k \sin^2 \theta}}, \qquad \mathbf{E}(\eta,k) = \int_0^{\eta} \sqrt{1- k \sin^2 \theta} \, \, d\theta.
\end{align}
Note that, while the zero temperature GCP is continuous everywhere, it is not differentiable at $m = \mu/X$ and $m = \mu$. The leading small temperature corrections to the zero temperature GCP can be obtained by applying Laplace's saddle point method to \eqref{eq:renormalizedGCP}:
\begin{align} \label{eq:smallTGCP}
	& \Psi_{\text{small } T} - \Psi_{T=0} \nonumber\\
	& =  - 
		\begin{cases}			
			\dfrac{\pi |\mu^2 - m^2Y^2|}{6 \beta^2 \sqrt{|\mu^2-m^2X^2|} \sqrt{|\mu^2-m^2|}}, \quad \text{if } m < \mu \text{ or } m > \dfrac{\mu}{X} \\[3ex]
			\dfrac{Y^2}{\pi\beta^2 X} e^{-\beta \mu} + \dfrac{e^{- \beta(m-\mu)} \sqrt{m}}{\sqrt{2 \pi \beta^3 }}\dfrac{1 - Y^2}{\sqrt{1-X^2}} + \dfrac{e^{- \beta(\mu - m X)} \sqrt{m}}{\sqrt{2\pi\beta^3}}\dfrac{Y^2 - X^2}{\sqrt{X(1-X^2)}}, \quad \text{if }  \dfrac{\mu}{X} \leq m \leq \mu .
		\end{cases} 
\end{align}
\hfill \rule{0.5\textwidth}{.2mm}
\end{strip}
Notice the presence of three exponentials at ``leading" order in the bottom case. Depending on the value of the minimizing $m$ one of these will dominate. 

The minimization here is a tedious process, but the outcome is straightforward. \cite{bdt} provides the expression for the GCP in the medium $m$ case ($\mu \geq m \geq \mu/X$) as well as its minimum. It can be checked that the same minimum is the only minimum of the entire piece-wise GCP. The other piece in \eqref{eq:smallTGCP} is essentially a featureless function in this regards. The minimizing values of the parameters $m$ and $\tilde{\nu}$ for the crystal phase, denoted $m_0$ and $\tilde{\nu}_0$, as implicit functions of $\mu$, are
\begin{align}
	&\mu = \dfrac{2 \mathbf{E}(\tilde{\nu}_0)}{\pi \sqrt{\tilde{\nu}_0}} - \sqrt{\dfrac{2}{\pi \beta}}   \dfrac{(1 -\tilde{\nu}_0)\mathbf{K}(\tilde{\nu}_0)}{\tilde{\nu}^{3/4}_0}\, e^{- \frac{\beta(\pi-2\mathbf{E}(\tilde{\nu}_0))}{\pi \sqrt{\tilde{\nu}_0}}}, \nonumber \\
	&m_0 = \dfrac{1}{\sqrt{\tilde{\nu}_0}}  - \sqrt{\dfrac{2}{\pi \beta}}   \dfrac{1}{\tilde{\nu}^{3/4}_0}\, e^{- \frac{\beta(\pi-2\mathbf{E}(\tilde{\nu}_0))}{\pi \sqrt{\tilde{\nu}_0}}},
\end{align}
while for the homogeneous massive phase both parameters take the constant value of one. 
The boundary between the two phases is simply $\mu = 2/\pi$; the leading exponential correction vanishes at the boundary ($\nu =1$). This is the same as the zero temperature critical chemical potential, and so the phase boundary at small temperatures is perpendicular to the $\mu$-axis. The minimized GCP is
\begin{equation} \begin{aligned}
	& \Psi^\text{min}_{\text{small } T} = \\
	&\begin{cases}
			\begin{aligned}
				- \dfrac{1}{4\pi}& - \dfrac{e^{- \beta(1-\mu)}}{\sqrt{2 \pi \beta^3 }} \\  &\text{(homogeneous massive phase)}
			\end{aligned} & \text{if } \mu \leq \dfrac{2}{\pi} \\[6ex]
			\begin{aligned}
				\dfrac{1}{4\pi} - &\dfrac{1}{2 \pi \tilde{\nu}_0} + \dfrac{1 - \tilde{\nu}_0}{2\tilde{\nu}_0^2} \dfrac{f_2(\tilde{\nu}_0)}{\sqrt{2 \pi \beta^3 }}\, e^{- \frac{\beta(\pi-2\mathbf{E}(\tilde{\nu}_0))}{\pi \sqrt{\tilde{\nu}_0}}} \\
				&(\text{crystal phase})
			\end{aligned} & \text{if } \mu \geq \dfrac{2}{\pi}
		\end{cases} 
\end{aligned}\end{equation}
At exactly zero temperature the exponentially small terms vanish, implying that the minimized zero temperature GCP is independent of the chemical potential below $\mu = 2/\pi$. This is the Silver Blaze phenomenon \cite{cohen2003}.

%%=============================================================%%
\subsubsection*{Ginzburg-Landau expansion at zero temperature}
%%=============================================================%%
Out of the three pieces of the zero-temperature GCP only one permits a small $m$ (or small $\lambda$) expansion, it being the bottom case in \eqref{eq:zeroTGCP}. The other two pieces explicitly require a non-zero $m$. The expansion is 
\begin{align}
		\Psi^{(N)}_{T=0} &= -\dfrac{\mu^2}{2\pi} + \dfrac{1}{2\pi}\lambda^2 \ln (2\mu) f_2(\nu) \nonumber \\
		& - \dfrac{2 \mu^2}{\pi} \sum_{n=2}^N  \dfrac{(2n-3)!}{n!(n-1)!} f_{2n}(\nu) \left(\dfrac{\lambda}{2\mu}\right)^{2n}
\end{align}
and it can be verified that this is identical to the zero temperature limit of \eqref{eq:GLExpAlln}. As expected, this expansion explicitly predicts that there is no crystal phase at zero temperature, confirming the result in Section \ref{sec4}. There is a single phase - the massless homogeneous phase - and thus no phase transitions.

The reason for this behaviour is now clear. The minimum of the exact zero temperature GCP lies not in the $m < \mu$ region, but in the $\mu < m < \mu/X$ region where a small $m$ expansion is not valid. Thus the Ginzburg-Landau approach was bound to fail from the start. The fact that the failure is caused by the piece-wise nature of the GCP also suggests a possible connection to the Silver Blaze phenomenon, but at the level of this manuscript the connection is purely algebraic.

%%=============================================================%%
%%=============================================================%%
\section{Conclusion}
%%=============================================================%%
%%=============================================================%%
The phase diagram of the Gross Neveu model in 1+1 dimensions was studied in detail using the Ginzburg-Landau approach. The approach has the great advantage of being generatable directly from the microscopic theory, and this fact holds in higher dimensions as well. It shows great success in predicting several important features of the phase diagram even at low order.

Increasing the order of the expansion improves the accuracy of the crystal phase except at zero temperature, where the expansion, regardless of its order, completely fails to predict the correct phase. This was proven by studying the asymptotic behaviour of the crystal phase. This behaviour is confirmed by the explicit zero temperature Ginzburg-Landau expansion explicitly constructed from the zero temperature grand canonical potential. Algebraically, this behaviour and the Silver-Blaze phenomenon seem to stem from the same source. Further investigation is required to establish a direct connection.

As the zero temperature Ginzburg-Landau expansion and the observable (or minimized) grand canonical potential are both obtained from the same finite temperature expression for the grand canonical potential, extraction of zero temperature phases from the Ginzburg-Landau expansion using resummation techniques or the theory of resurgence seems to be a promising future direction. This line of inquiry, if successful, could be used to probe zero temperature phases in a variety of models as a Ginzburg-Landau expansion is far easier to generate than the exact grand canonical potential.

%%=============================================================================%%
%%=============================================================================%%
%%=============================================================================%%
\begin{appendices}
%%=============================================================================%%
%%=============================================================================%%
%%=============================================================================%%

%%=============================================================================%%
%%=============================================================================%%
\section{Generating the Ginzburg-Landau expansion from GCP} \label{secA1}
%%=============================================================================%%
%%=============================================================================%%

The expansion is obtained by expanding the derivative of the dispersion \eqref{eq:dispersionDerivative} in powers of $m$ or $\lambda$:
\begin{align}
		\left| \dfrac{dq}{dE} \right| &= \sum_{n=0}^\infty \begin{pmatrix}
			2n \\ n \end{pmatrix} \dfrac{f_{2n}(\nu)}{(1+\sqrt{\nu})^{2n}} \left(\dfrac{m}{E}\right)^{2n} \nonumber \\
		&= \sum_{n=0}^\infty \begin{pmatrix} 2n \\ n \end{pmatrix} f_{2n}(\nu) \left(\dfrac{\lambda}{2E}\right)^{2n}.
\end{align}
and then substituting the above into \eqref{eq:renormalizedGCP} and carryinng out the integrals order by order. The functions $f_{2n}$ can be computed using
\begin{equation}
	f_{2n}(\nu) =   \begin{pmatrix} 2n \\ n \end{pmatrix}^{-1} (1+\sqrt{\nu})^{2n} \left(c_n(\nu) - Y^2 c_{n-1}(\nu) \right)
\end{equation}
where $c_n(\nu)$ obey the recursion formula \cite{thies2004}
\begin{equation}\begin{aligned}
		c_n =& \, \dfrac{1}{(1 + \sqrt{\nu})^2\, n} \Big((2 n -1 ) (1 + \nu)\, c_{n-1} \\
		& - (n - 1) (1-\sqrt{\nu})^2 \, c_{n-2} \Big), \quad c_0 = 1.
\end{aligned}\end{equation}
Expressions for a few low order functions are
\begin{align}
		&f_2(\nu) = 1 - \frac{\mathbf{E}}{\mathbf{K}}, \nonumber \\
		&f_4(\nu) = \dfrac{1}{3} \bigg(1 + 2\nu - (1 + \nu)\frac{\mathbf{E}}{\mathbf{K}}\bigg), \nonumber	\\
		&f_6(\nu) = \dfrac{1}{10} \bigg( (1 + 6\nu + 3\nu^2) - (1 + 4\nu + \nu^2) \frac{\mathbf{E}}{\mathbf{K}} \bigg), \nonumber \\
		&f_8(\nu) = \dfrac{1}{35} \bigg( (1 + 12\nu + 18 \nu^2 + 4\nu^3) \\
		 & \qquad \qquad - (1 + 9\nu + 9\nu^2 + \nu^3) \frac{\mathbf{E}}{\mathbf{K}} \bigg),  \nonumber \\
		&f_{10}(\nu) = \dfrac{1}{126} \bigg( (1 + 20\nu + 60 \nu^2 + 40\nu^3 + 5 \nu^4), \nonumber \\
		& \qquad \qquad - (1 + 16\nu + 36\nu^2 + 16\nu^3 + \nu^4) \frac{\mathbf{E}}{\mathbf{K}} \bigg). \nonumber
\end{align}

%%=============================================================================%%
%%=============================================================================%%
\section{Large order scalings} \label{secA2}
%%=============================================================================%%
%%=============================================================================%%
The scalings are obtained by minimizing the $N$-th order Ginzburg-Landau expansion for large $\mu$. The minimization conditions are
\begin{align} \label{eq:MinCoupled}
	&\partial_\lambda \Psi^{(N)} = 0 \implies \sum_{n=1}^N n \alpha_{2n} f_{2n} (\nu) \lambda^{2n-2} = 0, \nonumber \\
	&\partial_\nu \Psi^{(N)} = 0 \implies \sum_{n=1}^N \alpha_{2n} f'_{2n}(\nu) \lambda^{2n-2} = 0.
\end{align}
Eliminating $\alpha_2$ from these equations results in
\begin{equation}	\label{eq:MinCondAnyN}
	\sum_{n=2}^N \dfrac{(-1)^{n} g_n(\nu)} {n!(n-1)!} R_{2n-2}\left(\frac{\mu}{T}\right)\, \left(\dfrac{\lambda}{4 \pi T}\right)^{2n-2} = 0
\end{equation}
where $g_n(\nu) = n f'_2(\nu) f_{2n}(\nu) - f_2(\nu)  f'_{2n}(\nu)$ is a monotonic increasing non-negative function for all $n$. From this equation it is clear that $\lambda/T$ can depend on temperature and chemical potential only through the ratio $\mu/T$. 

In the vicinity of the line $L_N$ (equation $T = \mu/c_N$) the functions appearing in the minimization conditions behave as
\begin{align}
		& \ln (4 \pi T) \approx \ln \dfrac{4 \pi \mu}{c_N} + \epsilon_N(\mu), \nonumber\\
		& R_n\left(\frac{\mu}{T}\right) \approx  R_{n}(c_N) + \dfrac{c_N\, \epsilon_N(\mu)}{\pi} I_{n+1}(c_N), \, n \neq 2N-2 \nonumber \\
		& R_{2N-2}\left(\frac{\mu}{T}\right) \approx  \dfrac{c_N\, \epsilon_N(\mu)}{\pi} I_{2N-1}(c_N)
\end{align}
assuming a small deviation $\epsilon_N(\mu)$ \eqref{eq:NthOrderDeviation}. Then, in the vicinity of $L_N$, the only physical solution to \eqref{eq:MinCondAnyN} is 
\begin{align} \label{eq:largeMuM}
		\dfrac{\lambda^2}{T^2} &\approx \dfrac{16 \pi^2 N(N-1)\,g_{N-1} (\nu)\, R_{2N-4} (\mu/T)}{g_N(\nu) \, R_{2N-2} (\mu/T)} \nonumber\\
		& \propto \dfrac{1}{\epsilon_N(\mu)}.
\end{align}
This solution is always positive (required for a real $\lambda$) if $\epsilon_N(\mu)$ is positive while all the other roots of the polynomial equation \eqref{eq:MinCondAnyN} are negative (resulting in an imaginary $\lambda$). This behaviour follows from the fact that $R_{2N-2j}$, for any integer $j$ smaller than $N$, evaluated in the vicinity of the roots of $R_{2N}$ have the same sign as $\epsilon_N(\mu)$. While we have not attempted an analytic proof of this statement, it is a very simple matter to check it numerically.

Now, instead of eliminating $\alpha_2$ from the minimization equations \eqref{eq:MinCoupled}, one can eliminate $\alpha_{2N}$ to obtain
\begin{equation}	\label{eq:MinCondAnyNWithoutAlpha2N}
	\sum_{n=1}^{N-1} \left(\dfrac{n f_{2n}}{N f_{2N}} - \dfrac{f_{2n}'}{f_{2N}'} \right) \alpha_{2n}(T,\mu) \lambda^{2n-2} = 0.
\end{equation}
Combining this with \eqref{eq:largeMuM} and working at leading order in $\epsilon_N(\mu)$ gives the scaling
\begin{equation} \label{eq:epsilonScaling}
	\epsilon_N(\mu) \propto \dfrac{1}{N} \left(\frac{1}{\ln \mu}\right)^{\frac{1}{N-2}}.
\end{equation}



%\section{Example of another appendix section}\label{secA2}%
%Appendices may be used for helpful, supporting or essential material that would otherwise 
%clutter, break up or be distracting to the text. Appendices can consist of sections, %figures, tables and equations etc.

\end{appendices}
\newpage

\bibliographystyle{sn-mathphys}
\bibliography{gngl-bibliography}
%\bibliographystyle{unsrt}

\end{document}